\documentclass[conference]{IEEEtran}
\IEEEoverridecommandlockouts
\usepackage{cite}
\usepackage{amsmath,amssymb,amsfonts}
\usepackage{algorithmic}
\usepackage{graphicx}
\usepackage{textcomp}
\usepackage{xcolor}
\def\BibTeX{{\rm B\kern-.05em{\sc i\kern-.025em b}\kern-.08em
    T\kern-.1667em\lower.7ex\hbox{E}\kern-.125emX}}
\begin{document}

\title{The Data Science Fire Next Time: Innovative strategies for mentoring in data science\\
{\footnotesize %\textsuperscript{*}Note: Sub-titles are not captured in Xplore and
%should not be used
}
\thanks{BPDM was funded in part by a 2019 Google's eCSR Program grant, and grant funding from Facebook, Capital One, Drexel University, and Howard University}
}

\author{\IEEEauthorblockN{ Latifa Jackson}
\IEEEauthorblockA{\textit{Dept. of Pediatrics and Child Health} \\
\textit{Howard University}\\
Washington, DC USA \\
latifa.jackson@howard.edu}
\and
\IEEEauthorblockN{ Heriberto Acosta Maestre}
\IEEEauthorblockA{\textit{College of Computing and Engineering} \\
\textit{Nova Southeastern University}\\
Fort Lauderdale, FL USA \\
ha221@mynsu.nova.edu}
%\and
}
\maketitle

\begin{abstract}
As data mining research and applications continue to expand in to a variety of fields such as medicine, finance, security, etc., the need for talented and diverse individuals is clearly felt. This is particularly the case as Big Data initiatives have taken off in the federal, private and academic sectors, providing a wealth of opportunities, nationally and internationally. The Broadening Participation in Data Mining (BPDM) workshop was created more than 7 years ago with the goal of fostering mentorship, guidance, and connections for minority and underrepresented groups in the data science and machine learning community, while also enriching technical aptitude and exposure for a group of talented students. To date it has impacted the lives of more than 330 underrepresented trainees in data science. We provide a venue to connect talented students with innovative researchers in industry, academia, professional societies, and government. Our mission is to facilitate meaningful, lasting relationships between BPDM participants to ultimately increase diversity in data mining. This most recent workshop took place at Howard University in Washington, DC in February 2019. Here we report on the mentoring strategies that we undertook at the 2019 BPDM and how those were received.
\end{abstract}


\section{Introduction}
There is an acknowledged lack of diverse voices at the table in the field of data science. This lack of diversity has been identified as an educational ‘pipeline’ problem\cite{Cullinane2009STEMPipeline}, a retention problem\cite{Dunbar2019CSMotivation} and an interest problem\cite{Alvarez2009POCMentors}. These theories about what underlies the low levels of participation of women, ethnic and ability minorities in data science are important conversations to have in terms of identifying areas for significant improvement. These strategies for broadening participation are forward looking in terms of their prospective impact in the data science field. A significant investment has been made to expose students to research and industry careers in data science with the goal of ameliorating the diversity deficits\cite{Trujillo2015MentBenefits}. Our title calls back to James Baldwin's work \textit{The Fire Next Time}, where, like his work, we seek to identify mentoring recommendations that can better help guide the next generation of underrepresented trainees grow, thrive and contribute to computer science  disciplines.

In February 2019, the Broadening Participation in Data Mining (BPDM) workshop was held over a three day period (February 2-4, 2019) at Howard University, a historically black college/university (HBCU) located in Washington, DC. Since 2012, BPDM has developed a track record of supporting exceptional underrepresented undergraduate students, graduate students, and early career scientists in computer science, informatics, and data science related disciplines to participate in this workshop. Past participants have come from around the globe, with a strong focus on traditionally underrepresented minorities from the United States as defined by the US government\cite{Jackson2019Vulnerable}. BPDM recruited ten mentors in computer and data science from underrepresented groups including ethnic minorities, underrepresented gender and gender identity individuals, and underrepresented differently abled individuals\cite{Blaser2018Disability}.

One articulated challenge from past BPDMs that many underrepresented computer science students reported feeling is a sense of imposter syndrome. Imposter syndrome arises when individuals report feeling of inadequacy with feelings of lack of belonging and low sense of professional identity\cite{Parkman2016Imposter}. This paper represents an assessment of three innovative mentoring strategies that were employed during the BPDM  workshop to  encourage trainees to feel a sense of belonging within the workshop community and to gain trust for the purpose of encouraging the adoption of mentor advice.

\section{LITERATURE REVIEW}
Mentoring is a key contributor to the recruitment, retention, and productivity of underrepresented groups in Science, Technology, Engineering, and Mathematics(STEM)\cite{Huff2018Awareness}\cite{Stukes2018Trilogy}\cite{Solomon2018BlackWomen}. Each of these disciplines contributes to the domain areas of data science, the focus of our BPDM workshop. Traditional mentoring models involve the development of a technical and psychosocial relationship between an experienced and an inexperienced person\cite{Chesler2002WomenMentoring}. The lack of mentors specifically focused on mentoring underrepresented populations has been identified as increasingly important for the recruitment and retention of minority graduate students\cite{Thomas2007MentMin}\cite{Thomas2018Truth}. This deficit further weakens the networks that underrepresented minorities need to build their identities within the computer science domain. Indeed an assessment of undergraduate science students found a greater dependence on mentors in those who are retained in science than those who left the academic discipline\cite{Packard2004Retention}.

Our mentoring strategy uses the developed mentoring roadmap and network model based on mapping self-identified needs and career goals\cite{Montgomery2017MentoringRoadmap}. The mapping process includes an assessment of a personal need for mentoring to support successful advancement along a career roadmap. This method uses individual level mentoring to help mentees track their career course over a period of time. This approach has recently been used to track trainee well being in graduate education\cite{Woloshyn2019StudentWellbeing}. We extended this approach by using small group settings instead of individual constructs.

BPDM is a useful laboratory for the development of innovative strategies to build resilience in data science trainees whose career paths are directed toward computer science by addressing the way that our mentors interact with participants. BPDM draws underrepresented mentors from academic institutions, federal research centers, and industry partners. These mentors bring a diversity of experiences and perspectives with them to their mentoring experience. The goal this recruitment of mentors is to help participants whose daily professional exposure to other underrepresented groups is limited by their educational settings.

\section{METHODS}
In February 2019, BPDM workshop brought 55 participants and ten mentors together to Howard University at Washington DC. Participants were recruited from across the country (N= 25) and from computer science and related discipline students from Howard University, a historically Black College/University (N=30). The workshop was a collaborative effort between academic, industry, and federal government partners with the goal of introducing underrepresented students to a diversity of careers in data science, to connect students with mentors from underrepresented groups (women, ethnic minorities, and differently abled persons) who can help guide students in the pursuit of long term careers in data science, and  to enhance their network of peer data scientists who can support their careers along the way. BPDM worked collaboratively with Google’s Explore Computer Science Research for Undergraduate Women program to perform pre and post testing for BPDM participants to monitor programmatic quality improvement for future improvement\cite{Rorrer2019Challenges}. 

Workshop pretests (N=26 Individuals) consisted of 80 questions on participants sense of self efficacy, identity, belonging, and their sense of team belonging. While post-tests (N=17) consisted of 12 questions to evaluate the likelihood that participants would increase their desire to participate in computer science careers. These tests were administered online prior to participant arrival at the workshop and after participants returned to their home locations. Mentors were informally surveyed on the last day of the workshop to determine their perceived effectiveness of mentoring strategies.

\subsection{STRATEGY 1: Acknowledging Career Path Unpredictability}
A key component of mentoring work involves establishing the credentials of mentors as appropriate to offer advice to trainees and early career scientists. Traditional mentorship models build confidence in the professional acumen of the mentor, we sought to humanize mentors by asking them to portray their lived experiences as they progressed through their career. This was specifically accomplished by asking mentors to start their career stories with an acknowledgement that the place that you end up is not always the result of careful deliberation, but instead results from a series of opportunities that can appear stochastic if considered a forward in time walk. For many students, graduate programs are a very lonely experience full of hardships and obstacles they had never encountered nor feel they have the right skills to overcome. In these mentoring sessions, the participants realize that there are others who came through the same rocky path, who faced these obstacles, and were able to reach the goal.  

\subsection{STRATEGY 2: Identifying Career, Community and Personal Sacrifices}
One of the ways the career roadmap mentoring strategy was implemented was to ask identified mentors; both those who were from underrepresented groups and from majority groups; to identify what sacrifices they had to make in order to get to their current position. Past BPDM needs assessments showed that participants from underrepresented groups feel isolated because they are often asked to make difficult familial and economic sacrifices for their STEM education and career development. Even though we had recruited mentors from underrepresented groups, many of them indicated that they did not traditionally share their personal sacrifices made for their career. This was largely due to their desire to maintain a sense of professional distance with participants. One of the ways that this strategy was implemented was by placing participants into small groups where they stayed in the location and the mentors moved through the set of small groups to address the their experiences in a more intimate discussion environment. 

\subsection{STRATEGY 3: Deploying Mentors to Guide a 5 year Planning Path}  
Students were asked to track their career paths over the next five years across their next big career transition point such as undergraduate graduation, dissertation defense and transition into a career position. Workshop participants were divided into small groups based on their academic stage: undergraduate, early stage graduate student, Ph.D. candidates, postdoctoral fellows, and assistant professors. Each group had a senior mentor who served as advisors to the small student group as they discussed what they needed to do to move from their current career stage to their next career stage. Students identified barriers to their advancement, such as the Graduate Records Exam (GRE) cost, and both peer advice and mentor advice were solicited to built a coherent strategy to meet the identified participant group's challenges. Small groups then reported out their plans to the entire workshop, building a portrait of career planning that spanned from undergraduate education all the way through tenure in the academic pathway and into the first five years of the industry workforce.

\section{FINDINGS}
The BPDM workshop sought to use innovative mentoring strategies to increase self efficacy and participant sense of belonging in computer science. We employed a small group modification of the individual career mentoring strategy to increase the sense of belonging to an intellectual community, building rapport with senior mentors in the academic and industry arenas and exposing individuals to additional opportunities for career advancement.

When participants were asked about their sense of self efficacy prior (N=28, x=3.841, SD=0.74721, SE\textsubscript{Mean}=0.14121) to and after the workshop (N=16, x=3.9821, SD=0.87268, SE\textsubscript{Mean}=0.21817), we saw a significant change over the course of the workshop. This was also true for their sense of Identity prior (N=25, x=3.5067, SD=1.06009, SE\textsubscript{Mean}=0.21202) and after the workshop (N=16, x=3.75, SD=1.07841, SE\textsubscript{Mean}=0.2696), their sense of belonging prior to (N=26, x=3.7067, SD=1.04937, SE\textsubscript{Mean}=0.2058) and after the workshop, and finally their sense of team community prior to (N=26, x=4.3761, SD=0.57183, SE\textsubscript{Mean}=0.11215), and after (N=16, x=4.4792, SD=0.49644, SE\textsubscript{Mean}=0.12411) the BPDM workshop.

Interestingly, we found that actual student attitudes towards computer sciences (x\textsubscript{Pretest}=4.40 versus x\textsubscript{Posttest}= 4.35) did not significantly change over the course of the workshop, although we note that these mean scores were quite high in our pretest population (N=26, x\textsubscript{Pretest}= 4.40, SD= 0.52477, SE\textsubscript{Mean}= 0.10292) out of a possible 5 score. 

Anecdotally, we sought to collect qualitative data to assess the feelings of participants about their sense of community. One participants indicated that “this was by far the best opportunity I've ever had to engage in sustained, honest interaction with academic and industry professionals about research and career topics of interest to me.”  

Mentors described the sacrifices that they made for their careers included moving away from their families, making significant short term economic sacrifices in order to advance their computer science training, and finally having to successfully balance the relationship that they have with their families with their career responsibilities. For example, one industry mentor described their need to quit their full time job to attend a programming boot camp. Despite him saving money to accomplish this goal, his careful plan was upturned due to his roommate moving out and leaving him wholly responsible for previously shared living expenses. Several participants indicated that they appreciated that academic and industry mentors were willing to show the human costs of pursuing a career in computer science and they could see their own personal experiences reflected in the experiences of the mentors. These personal sacrifices are not often discussed in major conferences or by industry partners.

\section{DISCUSSION}
The Broadening Participation in Data Mining Workshop has been a multi-year mentoring endeavour to build the network and exposure of underrepresented participants to careers in data science, machine learning and computer science. This workshop sought to identify additional ways with with we can present a more realistic view of a career in data science whether in a research or industrial setting. Mentoring plays an important role in helping students from underrepresented groups\cite{Mejias2018CSPedagogy} reach their career potential. These students often are first generation undergraduate and graduate students who are facing new academic and career challenges while putting a brave face on these  seemingly insurmountable issues in order to reduce concern from their families\cite{Denner2018EngagingLatino}. They may feel additional pressure from family and close friends to achieve their academic goals as the vanguard of an educational and economic class switch, without the familial understandings of the sacrifices undertaken along the way.  

Over the past seven years, our organization has worked to build a community of over 300 past and present BPDM trainees who are contributing to computer and data science endeavours around the world. Here, We demonstrate that by using a modified roadmap mentoring strategy, participants improved their sense of self efficacy, belonging, identity, and their feelings of being part of a team during the workshop. In addition, participant attitudes to computer science were not significantly different before and after the workshop because participants self rated their enthusiasm for computer science and high. 

It is particularly gratifying to see significant evidence of increased reports of feelings of individual belonging and increased feeling of being part of a team. A sense of belonging has been noted to increase academic retention in geosciences careers\cite{Du2019WomenMent} for underrepresented women. This suggests that our mentoring strategy could contribute to retention efforts within computer and data science disciplines. In addition, work by Russell showed that including women and ethnic minorities in learning communities can help to increase and prolong their participation in STEM disciplines\cite{Russel2017LearningCommunities}.

A significant theme of our qualitative evaluation was the desire for relatable mentors across multiple dimensions. Mentors who provided students with a career path that they could follow were admired, but mentors who were willing to honestly address the difficulties inherent in undergraduate or graduate education were consistently referred to as providing more meaningful mentorship to participants. These mentors showed students that the training, economic, and personal obstacles are not impossible to overcome while concurrently helping participants manage their levels of frustration and expectations. 

Each year, student participants are encouraged to participate in the planning of the subsequent year's BPDM workshop to infuse the workshop with novel ideas garnered directly from trainees who are currently experiencing the rigors of computer or data science education. We also post opportunities for participants to engage in other mentoring opportunities such as conferences, training opportunities, internships, and workshops. These additional training opportunities allow participants to develop additional skills that can help to improve their engagement in this discipline.

Finally, we hope to continue to extend our mentoring studies to gain deeper insights relating to what mentoring activities can contribute to increasing the number of underrepresented groups who are able to make it to the computer science domain table and are able to be retained at that table. In particular we would like to evaluate the effect of our year round mentoring activities on feelings of well being within computer science domain areas.

\section*{Acknowledgments}

We thank the 2019 organizing committee of the Broadening Participation in Data Mining Workshop. We especially thank Orlando Ferrer and Caitlin Kuhlman for important discussions in the development of BPDM's mentoring plan. Finally we thank Audrey Rorrer and Breanna Spencer for their evaluation efforts as part of Google's Explore Computer Science Research Grant for Undergraduate Women.

\vspace{12pt}

\end{document}